\documentclass[letterpaper, 10 pt, conference]{ieeeconf}  

\IEEEoverridecommandlockouts                              

\usepackage{amsmath}
\usepackage{amssymb}
\usepackage{graphicx}
\usepackage{graphics}
\usepackage{subfigure}
\usepackage{tabularx}
\usepackage[ruled,linesnumbered]{algorithm2e}
\usepackage[colorlinks,urlcolor=black]{hyperref}
\usepackage{algorithmic}
\usepackage{caption}
\usepackage{wrapfig}
\usepackage{float}
\usepackage{subfigure}
\usepackage{hyperref}
\usepackage{cite}
\newtheorem{myDef}{Definition}

\newtheorem{mylem}{Lemma}
\newtheorem{myass}{Assumption}
\newtheorem{mycol}{Corollary}
\newtheorem{mypro}{Proposition}

\DeclareSymbolFont{symbolsC}{U}{txsyc}{m}{n}
\SetSymbolFont{symbolsC}{bold}{U}{txsyc}{bx}{n}
\DeclareMathSymbol{\Diamonddot}{\mathord}{symbolsC}{144}
\DeclareMathSymbol{\Boxdot}{\mathbin}{symbols}{237}

\title{ \bf
Distributed Monitoring of Robot Swarms with Swarm Signal Temporal Logic
}

\author{Ruixuan Yan and Agung Julius
\thanks{*This work was partially supported by the NSF through grant CNS-1618369. Both the authors are with the Department of Electrical, Computer, and
	Systems Engineering, Rensselaer Polytechnic Institute, Troy, NY 12180,
        {\tt\small Email: yanr5, julia2@rpi.edu}}%
}

\begin{document}

\maketitle
\begin{abstract}
In this paper, we develop a distributed monitoring framework for robot swarms so that the agents can monitor whether the executions of robot swarms satisfy Swarm Signal Temporal Logic (SwarmSTL) formulas. We define generalized moments (GMs) to represent swarm features. A dynamic generalized moments consensus algorithm (GMCA) with Kalman filter (KF) is proposed so that each agent can estimate the GMs. Also, we obtain an upper bound for the error between an agent's estimate and the actual GMs. This bound is independent of the motion of the agents. We also propose the rules for monitoring SwarmSTL temporal and logical operators. As a result, the agents can monitor whether the swarm satisfies SwarmSTL formulas with a certain confidence level using these rules and the bound of the estimation error. The distributed monitoring framework is applied to a swarm transporting supplies example, where we also show the efficacy of the Kalman filter in the dynamic generalized moments consensus process.
\end{abstract}

\section{INTRODUCTION}

A robot swarm is a multi-agent system composed of a large number of robots that can accomplish complicated tasks through interaction and cooperation \cite{brambilla2013swarm}. The issue of safety and correctness is very important for robot swarms. Temporal logic formulas are widely used to express such safety and correctness properties. The \textit{formal controller synthesis} part of the safety/correctness issue aims to construct control laws for robot swarms that result in executions satisfying temporal logic formulas \cite{kloetzer2007temporal,sahin2017provably,8250926,nikou2018timed}. The \textit{formal verification} part of the safety/correctness issue aims to check whether all the possible executions of robot swarms satisfy some temporal logic formulas \cite{DIXON20121429,10.5772/5769}. Most of the existing work performs formal verification of a robot swarm via formal verification of agents' executions. For example, Winfield et.al. \cite{10.5772/5769} build formal specifications for a swarm by combining the formal specifications of individual agents and determine whether the swarm specifications satisfy any emergent behavior. As an alternative to formal verification, \textit{temporal logic monitoring (TLM)} has been proposed, which aims to check whether finite executions of robot swarms satisfy temporal logic formulas.

When the size of a robot swarm is large, we cannot exhaustively perform TLM on the individual agents. Naturally, when we describe a swarm, we usually use abstract features (AF) of the swarm, such as the centroid, the size, or the shape of the swarm, while the behaviors of the agents
are less important. It is possible that each agent performs a distributed monitoring on the AF so that the agents know whether the AF satisfy some temporal logic formulas. Distributed monitoring on AF can improve the computational efficiency. Another advantage of distributed monitoring is that it can support the distributed control of robot swarms with temporal logic specifications. For example, if a robot swarm performs a task of supplies transportation, and it needs to drop the supplies only if its centroid reaches a certain region within 3 s. Distributed monitoring can help the agents decide whether to drop the supplies by monitoring whether the centroid of the swarm has reached the desired region within 3 s. 

In our previous paper \cite{yan2019swarm}, we proposed \textit{generalized moments (GMs)} to represent \textit{swarm features} and \textit{SwarmSTL} to analyze swarm behaviors. In this paper, we develop a distributed monitoring algorithm such that the agents can monitor the satisfaction of swarm features over SwarmSTL formulas. The swarm is considered as a graph composed of nodes and edges, where the nodes and edges represent agents and communication links, respectively. We assume the communication between agents is asynchronous, and the graph structure is time-invariant. We propose \textit{static and dynamic generalized moments consensus algorithms (GMCAs) with Kalman filter (KF)} so that the agents can estimate the GMs. These algorithms are based on a distributed average consensus algorithm (DACA) \cite{RGA}. There have been a lot of work on designing DACA \cite{spanos2005dynamic,zhu2010discrete,kia2019tutorial}, which requires the agents' estimates ($\zeta^j$) to track the average of the signals from individual agents ($\theta^j$). Most of the existing work assume a special initialization $\zeta^j(0)=\theta^j(0)$ or the agents' measurements are noiseless. In our work, we consider that the agents' measurements are noisy. We incorporate the KF into the GMCA so that each agent can estimate its state optimally and use its state estimate to perform the GMCA. This approach can get rid of the assumptions of the special initialization and the noiseless measurements. We show that if the agents remain static, their estimates of the GMs will converge to the actual GMs by performing the static GMCA with KF. For the dynamic GMCA with KF, we obtain an upper bound for the error between agents' estimates and the actual GMs. The result shows that this bound is independent of the agents' control inputs, i.e. the motion of the agents. This means the dynamic GMCA with KF can be performed simultaneously with other motion planning algorithms. 

We also propose SwarmSTL monitoring rules to help the agents compute the confidence level of swarm features satisfying temporal and logical operators. Using these rules and the estimation error bound from the dynamic GMCA with KF, agents can monitor whether the swarm features satisfy a SwarmSTL formula with a certain confidence level.

\textbf{Related Work.} Moarref et.al. \cite{8250926} synthesize decentralized controllers for a robot swarm with Linear Temporal Logic (LTL) specifications on both the swarm and individual levels. In \cite{kloetzer2009automatic}, the authors present a method to synthesize decentralized controllers for a robotic group with LTL specifications of visiting regions of interest. Kantaros et.al. \cite{8404066} propose a sampling-based method to synthesize optimal controllers for multi-agent systems with global temporal logic specifications. However, in this paper, we focus on the distributed monitoring of robot swarms with SwarmSTL specifications, not controller synthesis. Other work that is related to this paper is incorporating the average consensus algorithms into the distributed KF \cite{olfati2007distributed,8675461}, where the agents' measurements are coupled. Recently, He et.al. \cite{8675461} propose a distributed Kalman consensus filter for a multi-agent system with state equality constraints under time-based and event-triggered communication protocols. However, in this paper, each agent only has a noisy measurement of its own state and uses the KF to estimate its state.

\section{SWARM DYNAMICS AND SWARMSTL}\label{sec2}

\subsection{Dynamic Model and Features of A Swarm}

In this paper, the robot swarm works in a planar bounded environment $S\subset \mathbb{R}^2$. The discrete-time dynamic model of an agent is defined as
\begin{equation}
s(k+1) = s(k)+u(k)\label{ad},
\end{equation}
where $s\in S$ is the agent state (position), $k\in T$ is the time slot, $u\in U$ is the control input that directly controls the velocity, $T$ is the set of non-negative integers, and $U=\{u|\|u\|_\infty \leq u_{\max}\}$. Equivalently, we can write the state of an agent as $s = [s_x,s_y]^T$ and the control input as $u = [u_x, u_y]^T$. Let $N$ denote the number of agents in the swarm and $\mathbf{s}\in \mathbf{S}=S^N$ denote the state of a swarm, i.e. $\mathbf{s}=[{(s^1)}^T,{(s^2)}^T,...,{(s^N)}^T]^T$, where $s^j=[s_x^j,s_y^j]^T$ is the state of the $j$-th agent. The dynamic model of a swarm is defined as
\begin{equation}
\mathbf{s}(k+1)=\mathbf{s}(k)+\mathbf{u}(k),
\end{equation}
where $\mathbf{u}\in \mathbf{U}=U^N$, $\mathbf{u}=[{(u^1)}^T,{(u^2)}^T,...,{(u^N)}^T]^T$, and $u^j = [u^j_x,u^j_y]^T$ is the control input of the $j$-th agent. 

\begin{myDef}
	Let $P(s^j)$ denote a polynomial function of elements in $s^j$. We define the generalized moment (GM) $T^P:\mathbf{S}\rightarrow \mathbb{R}$ to represent a swarm feature, which is expressed as
	\begin{equation}
	T^P(\mathbf{s})= \frac{1}{N} \sum_{j=1}^NP(s^j).\label{gm}
	\end{equation}
\end{myDef}
For example, if $P_1(s^j) = s^j_x$, then we can define the mean of $x$ position of the agents as
$\bar{s}_{x} \triangleq T^{P_1}(\mathbf{s})=\frac{1}{N}\sum_{j=1}^{N}s^j_{x}.$
Similarly, for $P_2(s^j) = s^j_y$, we can define the mean of $y$ position of the agents as
$\bar{s}_{y} \triangleq T^{P_2}(\mathbf{s})=\frac{1}{N}\sum_{j=1}^{N}s^j_{y}.$
Here we focus on $v\in \mathbb{Z}_{> 0}$ generalized moments, and we denote these $v$ generalized moments as $\boldsymbol{\eta}=\mathbf{T^p(s)}=[\eta_1,\eta_2,\cdots,\eta_v]^T$, where $\mathbb{Z}_{> 0}$ is the set of positive integers.

\subsection{Swarm Signal Temporal Logic (SwarmSTL)}
The syntax of SwarmSTL is expressed as follows \cite{yan2019swarm}: 
\begin{equation}
\phi :=\top|\pi|\neg \phi|\phi_1 \wedge \phi_2|\phi_1 \vee \phi_2|\phi_1 \mathcal{U}_{[k_1,k_2]}\phi_2|\phi_1\mathcal{S}_{[k_1,k_2]}\phi_2|\epsilon,\label{syntax}
\end{equation}
where $\top$ is Boolean constant True, $\pi$ is an atomic proposition defined as $\pi:= \mathbf{a}^T\boldsymbol{\eta}\leq c$, $\neg, \wedge,\vee$ are Boolean operators representing ``negation", ``conjunction" and ``disjunction", respectively, $\mathcal{U}$ reads as ``\textit{Until}", $\mathcal{S}$ reads as ``\textit{Since}", $k_1,k_2\in T$, $\mathbf{a}=[a_1,...,a_v]^T\in \mathbb{R}^v$, $c\in \mathbb{R}$. We define $\epsilon\in \mathcal{E}$ as a logical proposition that represents an event, where $\mathcal{E}$ denotes all the possible events. The Boolean value of $\epsilon$ is known at any time slot, which is denoted by $b$. Since we do not use events in this paper, the detailed explanation of events is omitted.
Additionally, we define two useful temporal operators from $\mathcal{S}$: $\Diamonddot_{[k_1,k_2]}\phi=\top\mathcal{S}_{[k_1,k_2]}\phi$ (reads "eventually $\phi$ in the past") and $\boxdot_{[k_1,k_2]}\phi =\neg\Diamonddot_{[k_1,k_2]}\neg\phi$ (reads "always $\phi$ in the past"). We define ``$\Rightarrow$" as an implication operator, which is described as $\phi_1\Rightarrow \phi_2 := \neg\phi_1\vee \phi_2$.

The Boolean semantics of SwarmSTL can qualitatively measure the satisfaction of $(\boldsymbol{\eta},b)$ over $\phi$ at $k$, and $(\boldsymbol{\eta},b,k)\models\phi$ means $(\boldsymbol{\eta},b)$ satisfies $\phi$ at $k$. We denote the robustness degree of satisfaction of $(\boldsymbol{\eta},b)$ over $\phi$ at $k$ as $r(\boldsymbol{\eta},b,\phi,k)$. We claim that $(\boldsymbol{\eta},b,k)\models\phi$ if and only if $r(\boldsymbol{\eta},b,\phi,k)\geq 0$. We refer the readers to \cite{yan2019swarm} for more details about the Boolean semantics and robustness degree of satisfaction of SwarmSTL.

\section{PROBLEM STATEMENT AND APPROACH}\label{sec3}

In this paper, we consider an agent as a node and the communication links between agents as edges. We regard a robot swarm as a graph $G = \{D,E\}$, where $D$ is the set of nodes, and $E$ is the set of edges. For example, if agent $i$ can communicate with agent $j$, then $(i,j)\in E$. 
\begin{myass}\label{asm1}
	Assume the communication between agents is asynchronous, and each agent knows the graph structure that is time-invariant and has a noisy measurement of its own state. 
\end{myass}
The problem of distributed monitoring of swarm features with SwarmSTL can be reduced into the following problems:
 
\textbf{Problem 1:} Design an algorithm such that each agent can estimate the GMs;

\textbf{Problem 2:} Design a set of monitoring rules for SwarmSTL such that each agent can compute the confidence level of swarm features satisfying SwarmSTL formulas. 

From (\ref{gm}) we know that a generalized moment is the mean of some polynomial function $P(s^j)$. Hence we can adopt a DACA so that each agent can estimate the GMs. However, previous work on DACA \cite{RGA,spanos2005dynamic,zhu2010discrete} assume an agent's initial estimate $\zeta^j(0)$ is equal to its initial input signal $\theta^j(0)$, or that the agents' measurements are noiseless. Under these assumptions, if the agents' input signals are time-invariant, their estimates will converge to the average of $\theta^j(0)$ using a static DACA. If the agents measure their signals from the sensors that are noisy, then there will be a steady-state error for the above average consensus algorithms. We propose to incorporate the KF into the distributed GMs consensus process to solve \textbf{Problem 1}. Each agent can estimate its state optimally using the KF and use its state estimate to perform the distributed GMCA. We call this algorithm the GMCA with KF. If the agents remain static, then their estimates will converge to the actual GMs by performing a static GMCA with KF. We also propose a dynamic GMCA with KF so that each agent can estimate the GMs and compute an upper bound for the error between its estimate and the actual GMs. More details will be discussed in Section \ref{sec4}.

For \textbf{Problem 2}, we can use the estimation error bound, the robustness degree of an agent's estimate with respect to an atomic proposition $\pi$, and the Markov inequality, to compute an agent's confidence level of $\boldsymbol{\eta}$ satisfying $\pi$. We also define a set of rules to compute an agent's confidence level of the swarm features satisfying SwarmSTL temporal and logical operators, based on which an agent can compute the confidence level of the swarm features satisfying SwarmSTL formulas. More details will be discussed in Section \ref{sec5}.

\section{DISTRIBUTED GENERALIZED MOMENTS CONSENSUS WITH KALMAN FILTER}\label{sec4}

Considering the sensors are noisy, we adopt Kalman filter for each agent to estimate its state. For simplicity, we only discuss the consensus algorithm on one GM, and the same analysis can be applied to any GM.

\subsection{Optimal State Estimator}

The measurement model of an agent is defined as
\begin{equation}
y(k) = s(k) + v(k),
\end{equation}
where $y(k)$ is the measurement, $v(k)$ is the sensor noise. In this paper, we denote the trace of a matrix as $tr(\cdot)$.

\begin{myass}\label{ass2}
	Assume $v(k)$ follows a Gaussian distribution with $0$ mean and covariance matrix $K_{v_a}$ that is time-invariant, and each agent knows $E(\|v\|^2)$ is upper bounded by $v_{\max}$.
\end{myass}

The state estimate of a swarm is denoted as $\hat{\mathbf{s}}=[(\hat{s}^1)^T,(\hat{s}^2)^T,...,(\hat{s}^N)^T]^T$, where $\hat{s}^j=[\hat{s}^j_x,\hat{s}^j_y]^T $ is the state estimate of the $j$-th agent. The measurement model of a swarm is expressed as
\begin{equation}
\mathbf{y}(k) = \mathbf{s}(k)+\mathbf{v}(k),
\end{equation} 
where $\mathbf{y}=[(y^1)^T,(y^2)^T,...,(y^N)^T]^T, \mathbf{v}=[(v^1)^T,(v^2)^T,...,(v^N)^T]^T$, $y^j$ and $v^j$ are the measurement and the noise of the $j$-th agent, respectively.  
The covariance matrix of the state estimation error is $\Sigma=E[(\hat{\mathbf{s}}-\mathbf{s})(\hat{\mathbf{s}}-\mathbf{s})^T]$, and the covariance matrix of $\mathbf{v}$ is denoted as $K_v=E(\mathbf{v}\mathbf{v}^T)$, which is time-invariant. With Assumption \ref{ass2}, each agent knows $E(\|\mathbf{v}\|^2)$ is upper bounded by $Nv_{\max}$. 
\begin{myass}
	Assume each agent knows that the variance of the initial state estimation error is bounded, i.e. $E(\|\hat{s}^j(0)-s^j(0)\|^2)\leq s_{\max}$.
\end{myass}

The update of $\hat{\mathbf{s}}(k)$ and $\Sigma(k)$ is expressed as follows \cite{kalman1960new}:
\begin{equation}
\begin{split}
&K(k) = \Sigma(k-1)(\Sigma(k-1)+K_v)^{-1},\\
&\hat{\mathbf{s}}(k)= \hat{\mathbf{s}}(k-1)+\mathbf{u}(k-1)+K(k)(\mathbf{y}(k)-\\
&\hat{\mathbf{s}}(k-1)-\mathbf{u}(k-1)),\\
&\Sigma(k) = \left (I_N-K(k)\right)\Sigma(k-1)
\end{split}\label{kf}
\end{equation}
where $\hat{\mathbf{s}}(k)^- = E[\mathbf{s}(k)|\mathbf{Y}(k-1)]$, $\hat{\mathbf{s}}(k)=E[\mathbf{s}(k)|\mathbf{Y}(k)]$, $\mathbf{Y}(k)=[(\mathbf{y}(0))^T, (\mathbf{y}(1))^T,...,(\mathbf{y}(k))^T]^T$. 

\subsection{Static GMCA with Kalman Filter}\label{sgmckf}

For static GMCA with KF, the dynamic model of the swarm becomes 
\begin{equation}
	\mathbf{s}(k+1) = \mathbf{s}(k).
\end{equation} 
Hence the update of $\hat{\mathbf{s}}(k)$ and $\Sigma(k)$ is described as
\begin{equation}
\begin{split}
&\hat{\mathbf{s}}(k)= \hat{\mathbf{s}}(k-1)+K(k)\left(\mathbf{y}(k)-\hat{\mathbf{s}}(k-1)\right),\\
&\Sigma(k) = (I_N-\Sigma(k-1)(\Sigma(k-1)+K_v)^{-1})\Sigma(k-1).
\end{split}\label{kf_stc}
\end{equation}
The static GMCA with KF is inspired by the Randomized Gossip Algorithm \cite{RGA}. At $k=0$, each agent holds an initial estimate of the GM, i.e. $\zeta^j(0)=P(\hat{s}^j(0))$.  As the communication between agents is asynchronous, at each $k$, only two agents are communicating with each other. The probability of each agent being active is $\frac{1}{N}$. Here active means an agent can initiate communication with another agent. 
Let $W$ be an $N\times N$ matrix associated with the robot swarm, where entry $W_{ij}$ denotes the probability of agent $i$ communicating with agent $j$, and $\mathbf{1}\in \mathbb{R}^N$ be the vector of all ones. Let $\theta^j(k) =P(s^j(k))$, $\theta(k) = \mathbf{P}(\mathbf{s}(k)) = [\theta^1(k), ...,\theta^N(k)]^T$, $\hat{\theta}^j(k) =P(\hat{s}^j(k))$, $\hat{\theta}(k) = \mathbf{P}(\hat{\mathbf{s}}(k)) = [\hat{\theta}^1(k), ...,\hat{\theta}^N(k)]^T$, $\Delta\hat{\theta}(k) = \hat{\theta}(k)-\hat{\theta}(k-1)$, $\zeta(k)=[\zeta^1(k),\zeta^2(k),...,\zeta^N(k)]^T$, $\zeta(0) = \hat{\theta}(0)$, $\bar{\zeta}(0) = \frac{\mathbf{1}^T}{N}\zeta(0)$, $\Delta \bar{\hat{\theta}}(k) = \frac{\mathbf{1}^T}{N}\Delta \hat{\theta}(k)$, and $\bar{\eta}(k) = \bar{\theta}(k) = \frac{\mathbf{1}^T}{N}\theta(k)$. At $k+1$, if agent $i$ communicates with agent $j$, then they update their estimates with
\begin{equation}
\begin{split}
\zeta^i(k+1) = \frac{1}{2}\left(\zeta^i(k)+\zeta^j(k)\right)+\hat{\theta}^i(k+1)-\hat{\theta}^i(k),\\
\zeta^j(k+1) = \frac{1}{2}\left (\zeta^i(k)+\zeta^j(k)\right)+\hat{\theta}^j(k+1)-\hat{\theta}^j(k),
\end{split}\label{ijupdate}
\end{equation}
and the other agents update their estimates with
\begin{equation}
\zeta^p(k+1) = \zeta^p(k)+\hat{\theta}^p(k+1)-\hat{\theta}^p(k),\ (p\not = i,j).\label{pupdate}
\end{equation}
Suppose $I_N$ is the $N\times N$ identity matrix, $e^j=[0,...,1,...,0]^T\in\mathbb{R}^N$ with the $j$-th entry being 1 and all the other entries being 0. We can reformulate (\ref{ijupdate}) - (\ref{pupdate}) in the vector form as 
\begin{equation}
\zeta(k+1)=
V(k)\cdot 
\zeta(k)+\Delta\hat{\theta}(k)\label{up},
\end{equation}
where 
$V(k)$ has a probability of $\frac{1}{N}W_{ij}$ to be
$V_{ij}= I_N-(1/2)(e^i-e^j)(e^i-e^j)^T.$
We denote the expectation of $V(k)$ as $V$, and the estimation error at $k$ as $E(\|\zeta(k)-\bar{\eta}(k)\mathbf{1}\|_\infty)$, which can be decomposed as
\begin{equation}
\begin{split}
&E(\|\zeta(k)-\bar{\eta}(k)\mathbf{1}\|_\infty) = E(\|\zeta (k)-\bar{\hat{\theta}}(k)\mathbf
{1}+\\
&\bar{\hat{\theta}}(k)\mathbf{1}
-\bar{\eta}(k)\mathbf{1}\|_\infty),\\
&\leq E(\|\zeta (k)-\bar{\hat{\theta}}(k)\mathbf{1}\|_\infty)+E(\|\bar{\hat{\theta}}(k)\mathbf{1}-\bar{\eta}(k)\mathbf{1}\|_\infty),
\end{split}
\end{equation}
where $\bar{\hat{\theta}}(k) = \frac{\mathbf{1}^T}{N}\hat{\theta}(k)$. Hence we can prove the convergence of $E(\|\zeta(k)-\bar{\eta}(k)\mathbf{1}\|_\infty)$ via proving the convergence of $E(\|\zeta (k)-\bar{\hat{\theta}}(k)\mathbf{1}\|_\infty)$ and $E(\|\bar{\hat{\theta}}(k)\mathbf{1}-\bar{\eta}(k)\mathbf{1}\|_\infty)$. Let's first analyze $E(\|\bar{\hat{\theta}}(k)\mathbf{1}-\bar{\eta}(k)\mathbf{1}\|_\infty)$. 
Using the fact that a polynomial function $P(s^j)$ is locally Lipschitz on $S$ \cite{khalil2002nonlinear}, we know $T^P(\mathbf{s})$ is locally Lipschitz on $\mathbf{S}$. This implies there is a constant $\mathcal{L}_2\geq 0$ such that 
\begin{equation}
\begin{split}
&E(\|\bar{\hat{\theta}}(k)\mathbf{1}-\bar{\eta}(k)\mathbf{1}\|_\infty)=E(\|T^P\left(\hat{\mathbf{s}}(k)\right)-T^P\left(\mathbf{s}(k)\right)\|_\infty),\\
&\leq \mathcal{L}_2 E(\|\hat{\mathbf{s}}(k)-\mathbf{s}(k)\|_\infty),
\end{split}\label{kf_lip}
\end{equation}
where $E(\|\hat{\mathbf{s}}(k)-\mathbf{s}(k)\|_\infty) \leq \sqrt{tr(\Sigma(k))}$. Hence we can obtain an upper bound for $E(\|\bar{\hat{\theta}}(k)\mathbf{1}-\bar{\eta}(k)\mathbf{1}\|_\infty)$ by computing an upper bound for $tr(\Sigma (k))$. $\Sigma(k)$ can be reformulated as
$\Sigma(k)= K_v(\Sigma(k-1)+K_v)^{-1}\Sigma(k-1)
$, which leads to the following lemma.

\begin{mylem}\label{lem1}
	$\Sigma(k)$ follows the expression of $\Sigma(k) = K_v[K_v+k\Sigma(0)]^{-1}\Sigma(0)$, and the state estimation error $E(\|\hat{\mathbf{s}}(k)-\mathbf{s}(k)\|_\infty)\leq \sqrt{N^2s_{\max}v_{\max}/(v_{\max}+ks_{\max})}$ and converges to 0 asymptotically. 
\end{mylem}

\proof  We can prove Lemma \ref{lem1} by induction. At $k = 1$, we have $
\Sigma(1) = K_v(K_v+\Sigma(0))^{-1}\Sigma(0) = ((K_v+\Sigma(0))K_v^{-1})^{-1}\Sigma(0)
= (I+\Sigma(0)K_v^{-1})^{-1}\Sigma(0).$
Moreover, if $\Sigma(k) = K_v[K_v+k\Sigma(0)]^{-1}\Sigma(0) = (I+k\Sigma(0)K_v^{-1})^{-1}\Sigma(0)$ holds, we need to show $\Sigma(k+1) = K_v[K_v+(k+1)\Sigma(0)]^{-1}\Sigma(0)$ holds. This can be proved by
\begin{equation}
\begin{split}
&\Sigma(k+1) = K_v(\Sigma(k)+K_v)^{-1}(I+k\Sigma(0)K_v^{-1})^{-1}\Sigma(0),\\
& =K_v[(I+k\Sigma(0)K_v^{-1})\Sigma(k)+(K_v+k\Sigma(0))]^{-1}\Sigma(0),\\
& = K_v[(K_v+k\Sigma(0))K_v^{-1}\Sigma(k)+(K_v+k\Sigma(0))]^{-1}\Sigma(0),\\
& = K_v [(K_v+k\Sigma(0))((K_v+k\Sigma(0))^{-1}\Sigma(0)+I)]^{-1}\Sigma(0),\\
& = K_v\left[K_v+(k+1)\Sigma(0)\right]^{-1}\Sigma(0),\\
& = \left[\Sigma(0)^{-1}+(k+1)K_v^{-1}\right]^{-1}.
\end{split}
\end{equation}	
Therefore, $\Sigma(k)$ follows the expression of $\Sigma(k) = K_v[K_v+k\Sigma(0)]^{-1}\Sigma(0)$. Also, we can show that
	\begin{equation}
	\begin{split}
\lambda_1(\Sigma(k+1)) = \frac{1}{\lambda_N(\Sigma(0)^{-1}+(k+1)K_v^{-1})}\\
\leq \frac{1}{1/\lambda_1(\Sigma(0))+(k+1)/\lambda_1(K_v)},
\end{split}
\end{equation}	
which implies
	\begin{equation}
	\begin{split}
tr(\Sigma(k+1))\leq \frac{N^2s_{\max}v_{\max}}{v_{\max}+(k+1)s_{\max}}.
\end{split}
\end{equation}	
This proves $E(\|\hat{\mathbf{s}}(k)-\mathbf{s}(k)\|_\infty)\leq  \sqrt{N^2s_{\max}v_{\max}/(v_{\max}+ks_{\max})}$, and $E(\|\hat{\mathbf{s}}(k)-\mathbf{s}(k)\|_\infty)$ converges to 0 when $k\rightarrow \infty$.

\begin{mycol}\label{col1}
	A direct result of Lemma \ref{lem1} is $E(\|\bar{\hat{\theta}}(k)\mathbf{1}-\bar{\eta}(k)\mathbf{1}\|_\infty)\leq \mathcal{L}_2\sqrt{N^2s_{\max}v_{\max}/(v_{\max}+ks_{\max})}$ and $E(\|\bar{\hat{\theta}}(k)\mathbf{1}-\bar{\eta}(k)\mathbf{1}\|_\infty)$ converges to 0 asymptotically.
\end{mycol}
		
For simplicity, let $\delta_{\max}(k)=N^2s_{\max}v_{\max}/(v_{\max}+ks_{\max})$. The next task is to prove $E(\|\zeta (k)-\bar{\hat{\theta}}(k)\mathbf{1}\|_\infty)$ converges to 0. To prove this, we need the following proposition:
 
 \begin{mypro}\label{pro1}
\cite{RGA} 	The initial estimation error $E(\|\zeta(0)-\bar{\zeta}(0)\mathbf{1}\|^2)\leq E(\|\zeta(0)\|^2)$. Also, $E(\|V(0)\zeta(0) -\bar{\zeta}(0)\mathbf{1}\|^2)=E(\|V(0)(\zeta(0) -\bar{\zeta}(0)\mathbf{1})\|^2)\leq \lambda_2^2(V)\|\zeta(0)\|^2$, and $E(\|V(k-1)...V(0)\zeta(0)-\bar{\zeta}(0)\mathbf{1}\|^2)\leq \lambda_2^{2k}(V)\|\zeta(0)\|^2$. 
 \end{mypro}
 
Hereafter, we assume each agent knows $E(\|\zeta(0)-\bar{\zeta}(0)\mathbf{1}\|_\infty)\leq\zeta_{\max}$. Using the relationship that $E(\|\zeta (k)-\bar{\hat{\theta}}(k)\mathbf{1}\|_\infty)\leq E(\|\zeta (k)-\bar{\hat{\theta}}(k)\mathbf{1}\|)$, we have
\begin{equation}
\begin{split}
&E(\|\zeta (k)-\bar{\hat{\theta}}(k)\mathbf{1}\|_\infty)\leq 
E(\|V(k-1)...V(0)\zeta(0)+\\
&V(k-1)...V(1)\Delta \hat{\theta}(1)+...+\Delta \hat{\theta}(k)-[\bar{\hat{\theta}}(k)-\bar{\hat{\theta}}(k-1)\\
&+\bar{\hat{\theta}}(k-1)-\bar{\hat{\theta}}(k-2)...+\bar{\hat{\theta}}(1)-\bar{\zeta}(0)+\bar{\zeta}(0)]\mathbf{1}\|)\\
&\leq E(\|V(k-1)...V(0)\zeta(0)-\bar{\zeta}(0)\mathbf{1}+V(k-1)...V(1)\Delta \hat{\theta}(1)\\
&-\Delta \bar{\hat{\theta}}(1)\mathbf{1}+...+\Delta \hat{\theta}(k)-\Delta \bar{\hat{\theta}}(k)\mathbf{1}\|),\\
&\leq E(\|V(k-1)...V(0)\zeta(0)-\bar{\zeta}(0)\mathbf{1}\|)+E(\|V(k-1)...\\
&V(1)\Delta \hat{\theta}(1)-\Delta \bar{\hat{\theta}}(1)\mathbf{1}\|)+...+E(\|\Delta \hat{\theta}(k)-\Delta \bar{\hat{\theta}}(k)\mathbf{1}\|).
\end{split}\label{stakferr1}
\end{equation}
In order to prove $E(\|\zeta (k)-\bar{\hat{\theta}}(k)\mathbf{1}\|_\infty)$ converges to 0, we need to compute an upper bound for the one-step estimation error $E(\|V(k-1)\Delta \hat{\theta}(k-1)-\Delta\bar{\hat{\theta}}(k-1)\mathbf{1}\|)$. 

\begin{mylem}\label{lem2}
	The one-step estimation error satisfies $E(\|V(k-1)\Delta \hat{\theta}(k-1)-\Delta\bar{\hat{\theta}}(k-1)\mathbf{1}\|)\leq \lambda_2(V)\mathcal{L}_1\sqrt{\delta_{\max}(k-1)+\delta_{\max}(k-2)}$, where $\mathcal{L}_1\geq 0$. 
\end{mylem}

\proof Let $\mathfrak{e}_{k-1} = V(k-1)\Delta \hat{\theta}(k-1)-\Delta \bar{\hat{\theta}}(k-1)\mathbf{1}$. Using Jensen's inequality, we have the following inequality:
\begin{equation}
\begin{split}
&E(\|\mathfrak{e}_{k-1}\|) 
\leq\sqrt{E(\|\mathfrak{e}_{k-1}\|^2)},\\
&= \sqrt{\sum_{\delta_v=-\infty}^{\infty}\sum_{i,j=1}^N \delta_v^TV_{ij}^TV_{ij}\delta_v\frac{1}{N}W_{ij}}\\
&\sqrt{p[(\Delta \hat{\theta}(k-1)-\Delta \bar{\hat{\theta}}(k-1)\mathbf{1})=\delta_v]},\\
&= \sqrt{\sum_{\delta_v=-\infty}^{\infty} \delta_v^TV^TV\delta_vp[(\Delta \hat{\theta}(k-1)-\Delta \bar{\hat{\theta}}(k-1)\mathbf{1})=\delta_v]}.
\end{split}\label{exp_err}
\end{equation}
If we apply Proposition \ref{pro1} to $\delta_v^TV^TV\delta_v$, we can write (\ref{exp_err}) as
\begin{equation}
\begin{split}
&E(\|V(k-1)\Delta \hat{\theta}(k-1)-\Delta \bar{\hat{\theta}}(k-1)\mathbf{1}\|) \\
&\leq \lambda_2(V)\sqrt{E(\|\Delta \hat{\theta}(k-1)\|^2)},\label{kf_st1}
\end{split}
\end{equation}
where $p[(\Delta \hat{\theta}(k-1)-\Delta \bar{\hat{\theta}}(k-1)\mathbf{1})=\delta_v]$ denotes the probability of $\Delta \hat{\theta}(k-1)-\Delta \bar{\hat{\theta}}(k-1)\mathbf{1}=\delta_v$. 
As $\mathbf{P}(\mathbf{s})$ is locally Lipschitz on $\mathbf{S}$, $\exists \mathcal{L}_1\geq 0$ such that 
\begin{equation}
\begin{split}
\sqrt{E(\|\Delta \hat{\theta}(k)\|^2)} &= \sqrt{E(\|\mathbf{P}(\hat{\mathbf{s}}(k))-\mathbf{P}(\hat{\mathbf{s}}(k-1))\|^2)},\\
&\leq \mathcal{L}_1\sqrt{E(\|\hat{\mathbf{s}}(k)-\hat{\mathbf{s}}(k-1)\|^2)}.
\end{split}\label{kf_st2}
\end{equation}\\

In this subsection, both $\hat{\mathbf{s}}(k)$ and $\hat{\mathbf{s}}(k-1)$ are estimates of $\mathbf{s}(0)$, so $\sqrt{E(\|\hat{\mathbf{s}}(k)-\hat{\mathbf{s}}(k-1)\|^2)}$ can be written as
\begin{equation}
\begin{split}
&\sqrt{E(\|\hat{\mathbf{s}}(k)-\mathbf{s}(0)+\mathbf{s}(0)-\hat{\mathbf{s}}(k-1)\|^2)}\\
&\leq \sqrt{tr(\Sigma(k))+tr(\Sigma(k-1))}\leq \sqrt{\delta_{\max}(k)+\delta_{\max}(k-1)}.
\end{split}\label{stakf1}
\end{equation}
Hence we obtain $E(\|V(k-1)\Delta \hat{\theta}(k-1)-\Delta\bar{\hat{\theta}}(k-1)\mathbf{1}\|)\leq \lambda_2(V)\mathcal{L}_1\sqrt{\delta_{\max}(k-1)+\delta_{\max}(k-2)}$.

\begin{mycol}
	With Lemma \ref{lem2} and Proposition \ref{pro1}, we can rewrite (\ref{stakferr1}) as
	\begin{equation}
	\begin{split}
	&E(\|\zeta (k)-\bar{\hat{\theta}}(k)\mathbf{1}\|_\infty) \leq \lambda_2^k(V)\sqrt{N}\zeta_{\max}+\lambda_2^{k-1}(V)\mathcal{L}_1\\
	&	\sqrt{\delta_{\max}(1)+\delta_{\max}(0)}+...+\mathcal{L}_1\sqrt{\delta_{\max}(k)+\delta_{\max}(k-1)},\\
	&\!\leq\! \lambda_2^k(V)\!\sqrt{N}\zeta_{\max}\!+\!\mathcal{L}_1\!\!\sum_{k_t=1}^k\!\!\lambda_2^{k-k_t}(V)\sqrt{\delta_{\max}(k_t)\!+\!\delta_{\max}(k_t\!-\!1)}.
	\end{split}
	\end{equation}
Also, $E(\|\zeta (k)-\bar{\hat{\theta}}(k)\mathbf{1}\|_\infty)$ converges to 0 asymptotically. 
\end{mycol}

This is because when $k\rightarrow \infty$, $\delta_{\max}(k)$ and $\delta_{\max}(k-1)$ converge to 0. Also, $\lambda_2(V)<1$, so $\lambda_2^k(V)\rightarrow 0$ when $k\rightarrow \infty$. Combined with Corollary \ref{col1}, we can prove $E(\|\zeta(k)-\bar{\eta}(k)\mathbf{1}\|_\infty)$ converges to 0 asymptotically, which means each agent's estimate of the GM converges to the actual GM. The \textbf{static estimation error bound with KF} is denoted as
\begin{equation}
\begin{split}
&E(\|\zeta (k)-\bar{\eta}(k)\mathbf{1}\|_\infty) \leq \lambda_2^k(V)\sqrt{N}\zeta_{\max}+\mathcal{L}_1\sum_{k_t=1}^k\lambda_2^{k-k_t}(V)\\
&\sqrt{\delta_{\max}(k_t)+\delta_{\max}(k_t-1)}+\mathcal{L}_2\sqrt{\delta_{\max}(k)}.
\end{split}\label{st_kf_err}
\end{equation}

\subsection{Dynamic GMCA with Kalman Filter}\label{dgmkf}

For dynamic GMCA with KF, the update of $\hat{\mathbf{s}}(k)$ and $\Sigma(k)$ is expressed as (\ref{kf}), and our goal is to obtain an upper bound for $E(\|\zeta(k)-\bar{\eta}(k)\mathbf{1}\|_\infty)$. With (\ref{kf}), we could write $\hat{\mathbf{s}}(k)$ as
\begin{equation}
\begin{split}
&\hat{\mathbf{s}}(k) =  \hat{\mathbf{s}}(k-1)+\mathbf{u}(k-1)
+K(k)(\mathbf{s}(k-1)-\\
&\hat{\mathbf{s}}(k-1)+\mathbf{v}(k)),
\end{split}\label{dy_dsk}
\end{equation}
which leads to the following lemma.
\begin{mylem}\label{lem3}
	For dynamic GMCA with KF, the one-step estimation error satisfies 
	\begin{equation}
	\begin{split}
	&E(\|V(k-1)\Delta\hat{\theta}(k-1)-\Delta\bar{\hat{\theta}}(k-1)\mathbf{1}\|)\leq \lambda_2(V)\mathcal{L}_1\\
	&\sqrt{\delta_{\max}(k-1)+\delta_{\max}(k-2)+2Nu_{\max}^2}.
	\end{split}
	\end{equation}
\end{mylem}
\proof For dynamic GMCA with KF, (\ref{kf_st1}) and (\ref{kf_st2}) still hold. The upper bound for $\sqrt{E(\|\hat{\mathbf{s}}(k)-\hat{\mathbf{s}}(k-1)\|^2)}$ is
\begin{equation}
\begin{split}
&\sqrt{E(\|\hat{\mathbf{s}}(k)-\mathbf{s}(k)+\mathbf{s}(k-1)+\mathbf{u}(k-1)-\hat{\mathbf{s}}(k-1)\|^2)}\\
&\leq \sqrt{tr(\Sigma(k))+tr(\Sigma(k-1))+\|\mathbf{u}(k-1)\|^2},\\
&\leq \sqrt{\delta_{\max}(k)+\delta_{\max}(k-1)+2Nu_{\max}^2},
\end{split}
\end{equation}
hence we have $E(\|V(k-1)\Delta \hat{\theta}(k-1)-\Delta\bar{\hat{\theta}}(k-1)\mathbf{1}\|)\leq \lambda_2(V)\mathcal{L}_1\sqrt{\delta_{\max}(k-1)+\delta_{\max}(k-2)+2Nu_{\max}^2}$.
\begin{mycol}
	With Lemma \ref{lem3}, we obtain the \textbf{dynamic estimation error bound with KF (DBKF)} as follows:
	\begin{equation}
	\begin{split}
	&E(\|\zeta (k)-\bar{\eta}(k)\mathbf{1}\|_\infty) \leq \lambda_2^k(V)\sqrt{N}\zeta_{\max}+\mathcal{L}_1\sum_{k_t=1}^k\lambda_2^{k-k_t}(V)\\
	&\sqrt{\delta_{\max}(k_t)+\delta_{\max}(k_t-1)+2Nu_{\max}^2}+\mathcal{L}_2\sqrt{\delta_{\max}(k)}.
	\end{split}\label{dy_kf_bd}
	\end{equation}	
\end{mycol}

For simplicity, let $\rho(k)= \lambda_2^k(V)\sqrt{N}\zeta_{\max}+\mathcal{L}_1\sum_{k_t=1}^k\lambda_2^{k-k_t}(V)\sqrt{\delta_{\max}(k_t)+\delta_{\max}(k_t-1)+2Nu_{\max}^2}+\mathcal{L}_2\sqrt{\delta_{\max}(k)}$, which implies $\rho(k)$ is independent of $\mathbf{u}$. The above analysis is for one GM, and we can generalize the result to $v$ GMs. Let $\boldsymbol{\zeta}^j=[\zeta_1^j,...,\zeta_v^j]^T$ denote the $j$-th agent's estimate of $\boldsymbol{\eta}$, and $\rho_i(1\leq i\leq v)$ denote the bound of the estimation error of $\eta_i$. In Section \ref{sec5}, we will explain how to compute an agent's confidence level of swarm features satisfying SwarmSTL formulas using this \textbf{DBKF}. 

\subsection{Optimizing The Convergence Rate}

From (\ref{st_kf_err}) we know the convergence rate of the estimation error is controlled by $\lambda_2(V)$. The fastest convergence rate can be achieved by solving the following Semidefinite Programming (SDP) problem  \cite{RGA}:
\begin{equation}
\begin{aligned}
&\text{minimize}& &q,\\
&\text{subject to}& & W_{ij}\geq 0,\ W_{ij} = 0\ \text{if $(i,j)\not \in E$,}\\
&&&V = \frac{1}{N}\sum_{i,j=1}^N W_{ij}V_{ij},\ V-\frac{1}{N}\mathbf{1}\mathbf{1}^T\preceq qI_N,\\
&&&\sum_j W_{ij} = 1, \forall i,\\
\end{aligned}\label{sdp}
\end{equation}
where $V-\frac{1}{N}\mathbf{1}\mathbf{1}^T\preceq qI_N$ means $(qI_N-V+\frac{1}{N}\mathbf{1}\mathbf{1}^T)$ is positive semidefinite. With Assumption \ref{asm1}, the agents can perform a distributed optimization of $\lambda_2(V)$. In this paper, we use the SDP mode in the cvx\footnote{http://cvxr.com/cvx/doc/sdp.html} toolbox to solve (\ref{sdp}). 

\section{DISTRIBUTED MONITORING RULES}\label{sec5}

The \textbf{DBKF} can help us derive the confidence level of swarm features satisfying SwarmSTL formulas. Let $m^j(k)$ denote the robustness degree of $\boldsymbol{\zeta}^j$ over $\pi$ at $k$. The rules for distributed monitoring of swarm features satisfying SwarmSTL temporal and logical operators are described as follows. 

\begin{mylem}
The $j$-th agent's confidence level of $(\boldsymbol{\eta},b)$ satisfying $\phi$ at $k$ is described as follows:
\begin{equation}
\begin{aligned}
&Pr_j((\boldsymbol{\eta},b,k)\models \pi)
\geq 
\begin{cases}
1- \frac{1}{m^j(k)}\max\limits_{1\leq i\leq v}\{\rho_i(k)\},\ \\\text{if}\ m^j(k)>0,\\
0,\ \text{otherwise},
\end{cases}\\
&Pr_j((\boldsymbol{\eta},b,k)\models \phi_1\wedge\phi_2)\geq Pr_j((\boldsymbol{\eta},b,k)\models \phi_1)\\
&+Pr_j((\boldsymbol{\eta},b,k)\models \phi_2)-1,\\
&Pr_j((\boldsymbol{\eta},b,k)\models \phi_1\vee\phi_2)\geq 1-\min\{1-Pr_j((\boldsymbol{\eta},b,k)\\
& \models\phi_1),1-Pr_j((\boldsymbol{\eta},b,k)\models \phi_2)\},\\
&Pr_j((\boldsymbol{\eta},b,k)\models \phi_1\mathcal{S}_{[k_1,k_2]}\phi_2) \geq 1-\min_{k'\in[k-k_2,k-k_1]}\{1-\\
&Pr_j((\boldsymbol{\eta},b,k')\models \phi_2)+\sum_{k''=k'}^{k}1-Pr_j((\boldsymbol{\eta},b,k'')\models \phi_1)\},
\end{aligned}\label{cl_est}
\end{equation}
where $Pr_j((\boldsymbol{\eta},b,k)\models \phi)$ denotes the $j$-th agent's confidence level of $(\boldsymbol{\eta},b)$ satisfying $\phi$ at $k$.
\end{mylem}

\proof Let $\boldsymbol{\theta}_i = [\theta^1_i, ...,\theta^N_i]^T(1\leq i\leq v)$ and $\eta_i = \frac{\mathbf{1}^T}{N}\boldsymbol{\theta}_i$, where $\theta_i^j = P_i(s^j)$. Let $\zeta_i = [\zeta_i^1,...,\zeta_i^N]^T$ denote the agents' estimates of $\eta_i$. Using Markov Inequality, we know the estimation error of $\eta_i$ satisfies
$
Pr_j(\|\zeta_i(k)-\eta_i(k)\mathbf{1}\|_\infty\leq m^j(k))\geq 1-E(\|\zeta_i(k)-\eta_i(k)\mathbf{1}\|_\infty)/m^j(k)
\geq 1-\rho_i(k)/m^j(k).
$
For $v$ generalized moments, we could take the maximum of $\rho_i$ as the estimation error bound for $\boldsymbol{\eta}$. Hence for an atomic proposition $\pi$, the $j$-th agent's confidence level of $(\boldsymbol{\eta},b)$ satisfying $\pi$ at $k$ is $
Pr_j\left ((\boldsymbol{\eta},b,k)\models \pi\right)\geq
1- \max\limits_{1\leq i\leq v}\{\rho_i(k)\}/{m^j(k)}$ if $m^j(k)>0$. Otherwise, it is 0.
With this confidence level, it is also straightforward to show the $j$-th agent's confidence level of $(\boldsymbol{\eta},b)$ satisfying other operators at $k$. 

Using these monitoring rules of SwarmSTL temporal and logical operators, each agent can monitor the satisfaction of swarm features with respect to SwarmSTL formulas. 

\section{CASE STUDY} \label{sec6}

\begin{figure}[htbp]
	\centering
	\subfigure[Snapshot at $k = 11785$.]{
		\includegraphics[width=1.65in]{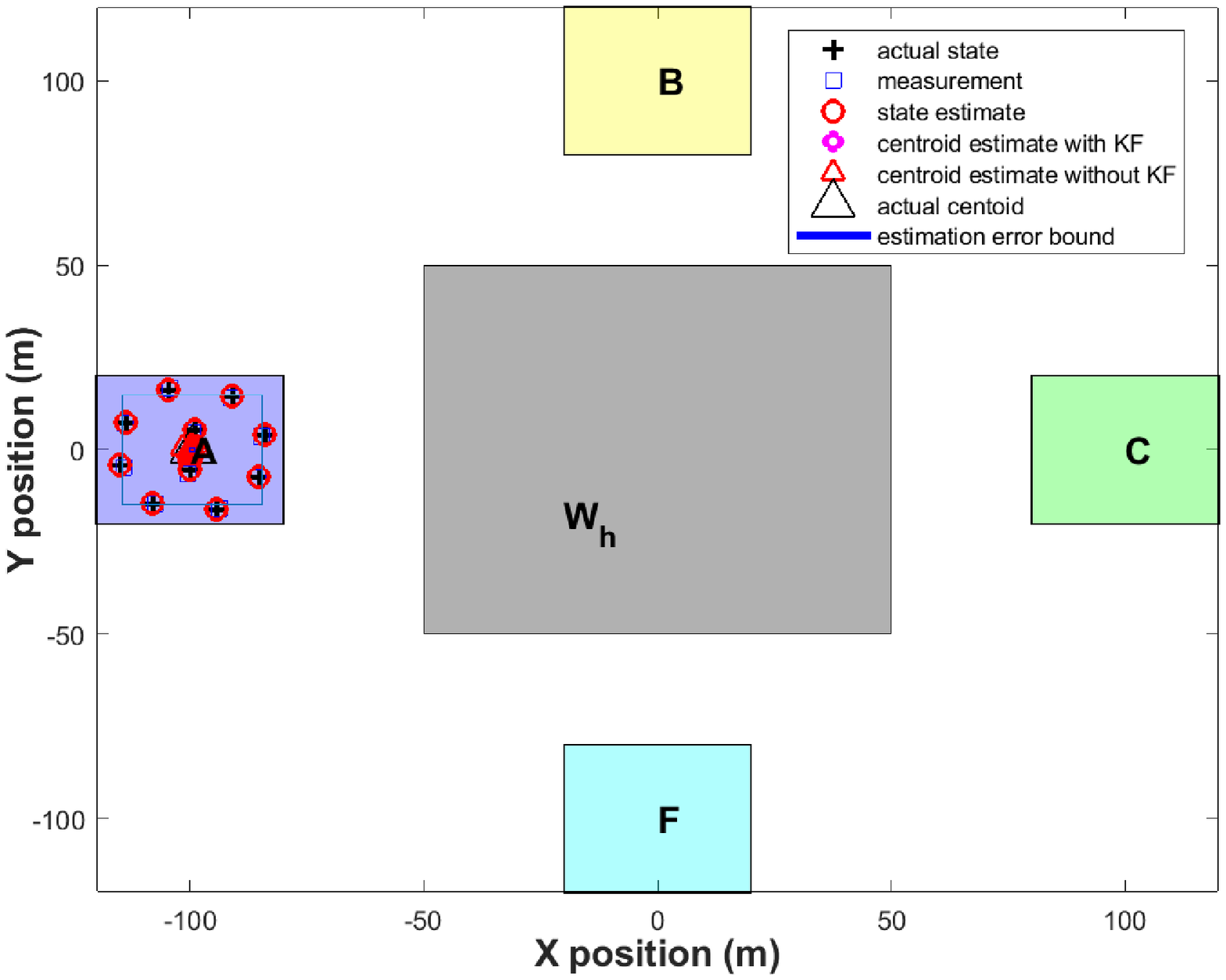}
	}\hspace{-10pt}
	\subfigure[Snapshot at $k = 31937$.]{
		\includegraphics[width=1.65in]{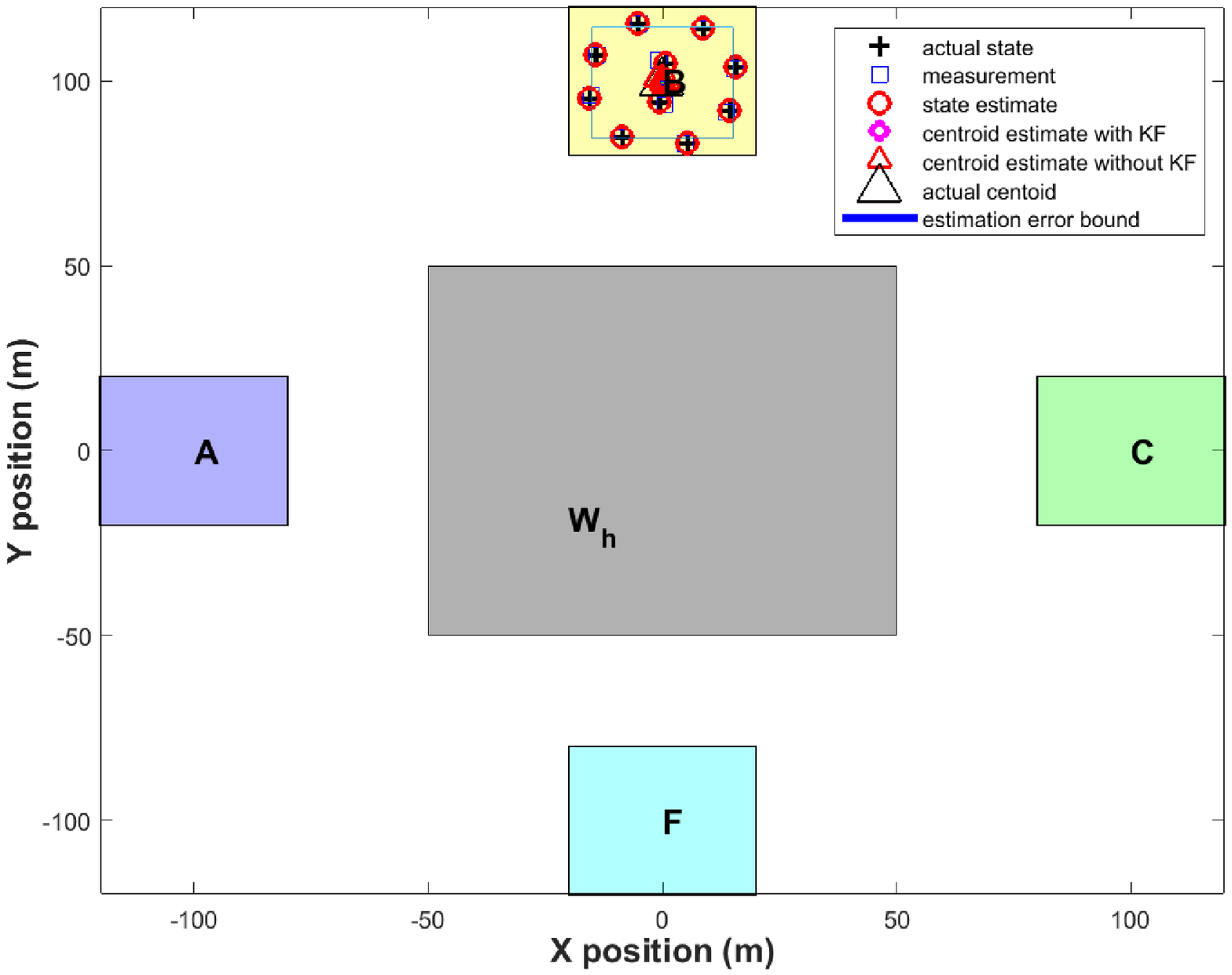}	
}\vspace{-4pt}
\quad
	\subfigure[Snapshot at $k = 51789$.]{
		\includegraphics[width=1.65in]{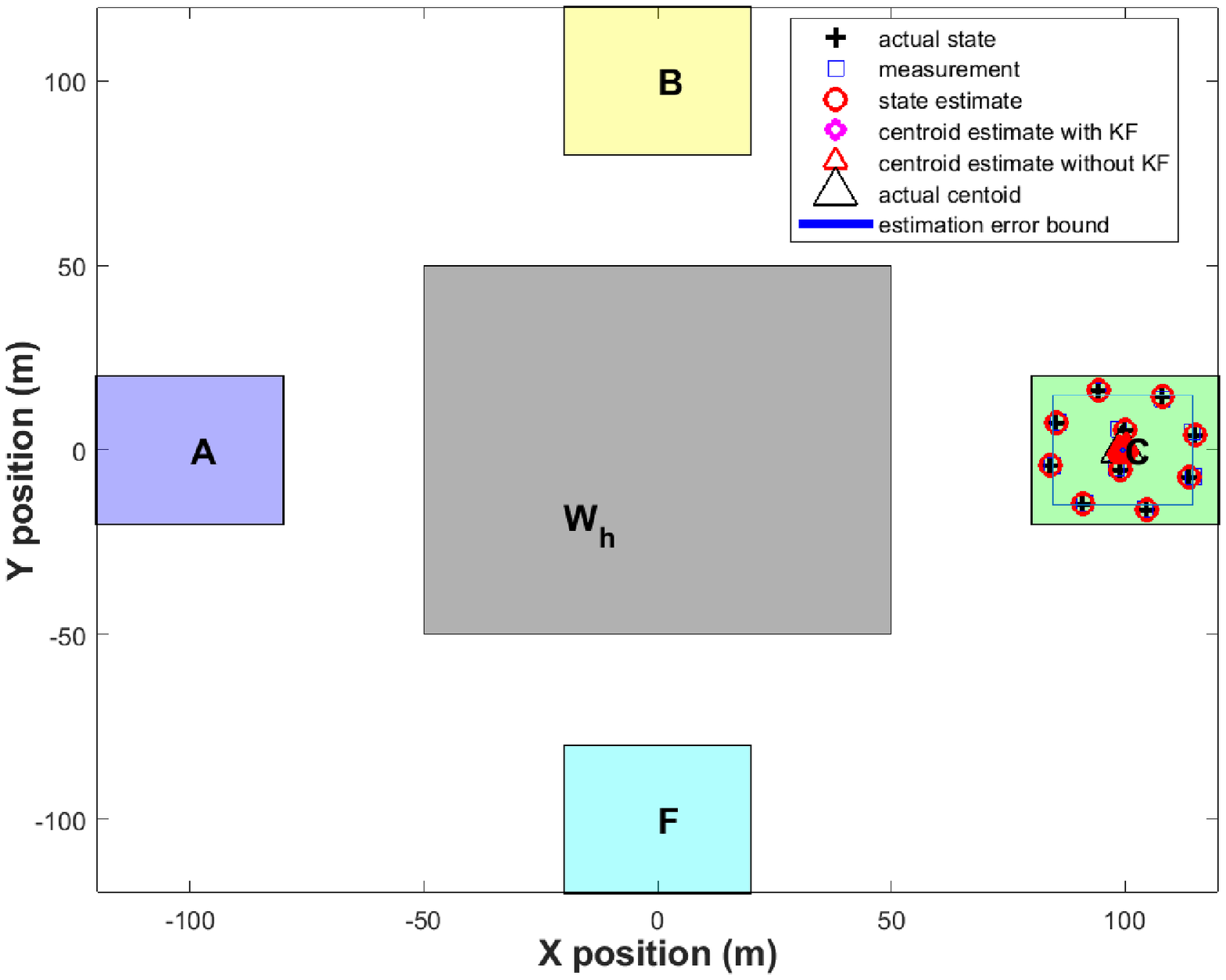}
	}\hspace{-11pt}
	\subfigure[Snapshot at $k = 71641$.]{
		\includegraphics[width=1.65in]{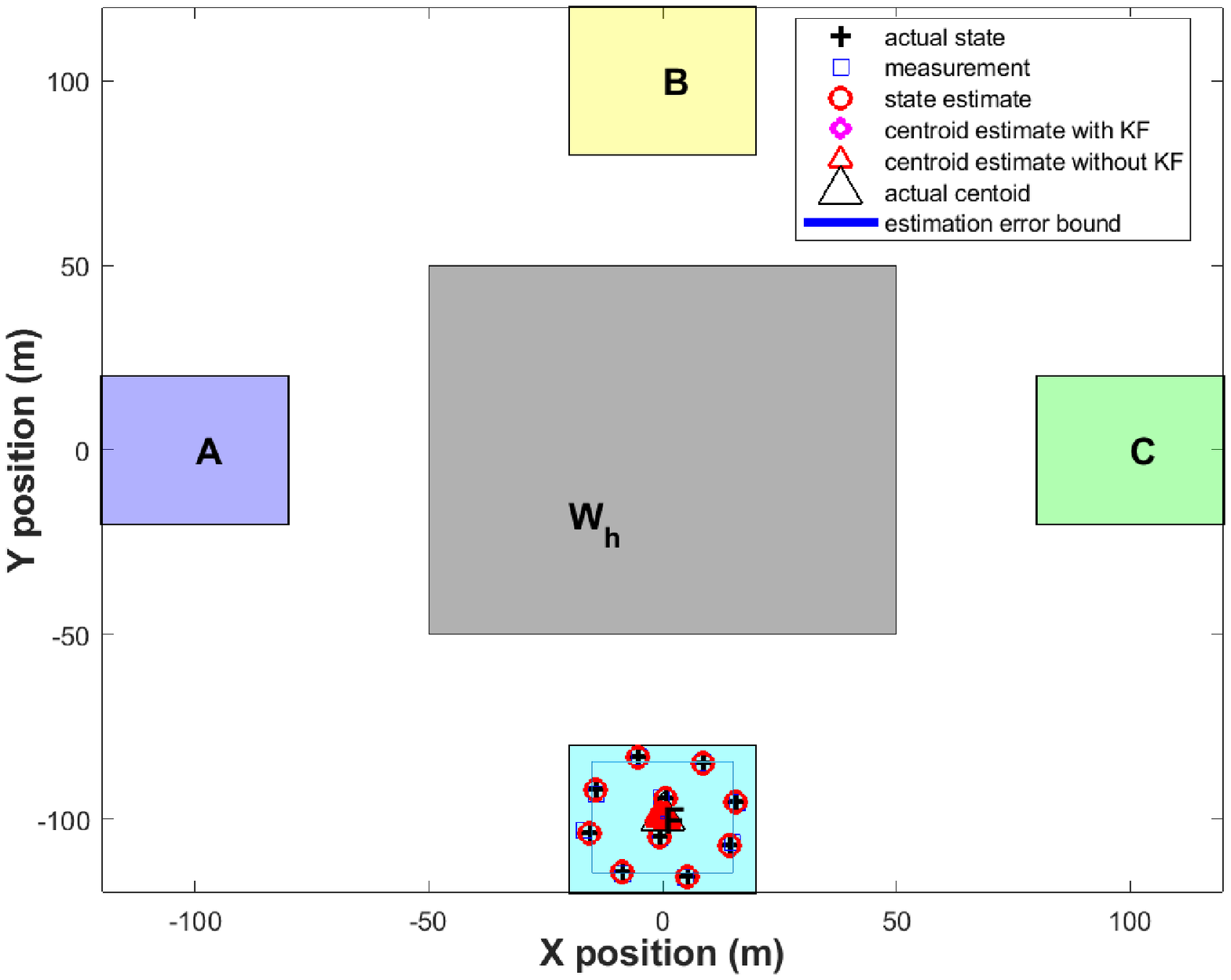}
	}
\vspace{-8pt}
	\caption{Snapshots of the simulation at different $k$.}
	\label{fig2}
\end{figure}

We now demonstrate the distributed monitoring of robot swarms with SwarmSTL formulas using a swarm transporting supplies example. Suppose there is a virus outbreak in four regions $A,B,C,F$, and the robot swarm needs to transport the supplies to these four regions back and forth. Specifically, the robot swarm first transports the supplies from $W_h$ to $A$, then goes back to $W_h$ to take another supply and transports it to $B$, then follows the same idea to transport the supplies to $C$ and $F$, i.e. the swarm follows a trajectory in the sequence of $W_h\rightarrow A \rightarrow W_h\rightarrow B\rightarrow W_h\rightarrow C\rightarrow W_h\rightarrow D\rightarrow W_h$. In the simulation, we set $N=10$, and we adopt a flocking model \cite{reynolds1987flocks} and a motion planning model so that the weight of the supplies can be distributed among the agents and the swarm can transport the supplies to the four regions. Some snapshots of the simulation are shown in Fig. \ref{fig2}. A video of the simulation is uploaded to \url{https://tinyurl.com/swarmstlmonitor}. 
As the robot swarm needs to transport the supplies from $W_h$ to the four regions back and forth, it cannot stay in the warehouse ($W_h$) for too long. Hence we propose a SwarmSTL formula $\phi_R := \Diamonddot_{[1000,2000]}\phi_{W_h}\Rightarrow \Diamonddot_{[0,800]}\neg \phi_{W_h}$ that specifies the robot swarm to stay in $W_h$ for at most $1000$ time slots, where $\phi_{W_h} := \bar{s}_x\geq -50\wedge \bar{s}_x\leq 50\wedge \bar{s}_y\geq -50\wedge \bar{s}_y\leq 50$. $\phi_R$ reads as if $\exists k'\in[1000,2000]$ such that $\phi_{W_h}$ is satisfied at $k-k'$, then $\exists k''\in[0,800]$ such that $\neg\phi_{W_h}$ is satisfied at $k-k''$. The agents need to perform the consensus on the centroid $\bar{s}=[\bar{s}_x,\bar{s}_y]^T$. At $k=0$, each agent sets $\boldsymbol{\zeta}^j(0) = [\zeta^j_x(0),\zeta^j_y(0)]^T = [\hat{s}^j_x(0),\hat{s}^j_y(0)]^T$, and computes the optimal $\lambda_2(V)$ by solving the SDP problem (\ref{sdp}). Let $\hat{\mathbf{s}}_x = [\hat{s}^1_x,...,\hat{s}^N_x]^T$, $\hat{\mathbf{s}}_y = [\hat{s}^1_y,...,\hat{s}^N_y]^T$, $\zeta_x = [\zeta_x^1,...,\zeta^N_x]^T$, and $\zeta_y = [\zeta_y^1,...,\zeta^N_y]^T$. At $k+1$, the agents perform the KF (\ref{kf}) to update $\hat{\mathbf{s}}_x$ and $\hat{\mathbf{s}}_y$, and update their estimates of the centroid by
\begin{equation}\\
\begin{split}
\zeta_x(k+1) = V(k)\zeta_x(k)+\hat{\mathbf{s}}_x(k+1)-\hat{\mathbf{s}}_x(k),\\
\zeta_y(k+1) = V(k)\zeta_y(k)+\hat{\mathbf{s}}_y(k+1)-\hat{\mathbf{s}}_y(k).
\end{split}
\end{equation}
In the meantime, the agents compute the bound of the centroid estimation error, $\rho_{\max} = \max\{\rho_x,\rho_y\}$, and the confidence level of $\bar{s}$ satisfying $\phi_R$, where $\rho_x,\rho_y$ are the estimation error bounds for $\bar{s}_x,\bar{s}_y$, respectively. The purpose of adopting the flocking model and the motion planning model is to show the agents can perform the centroid consensus and the flocking behavior, motion planning task simultaneously, because $\rho_{\max}$ is independent of $\mathbf{u}$. 

\begin{figure}
	\vspace{0.15cm}
	\centering
	\includegraphics[width = 2in]{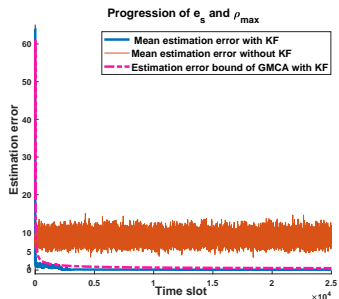}
		\vspace{-8.5pt}
	\caption{Progression of $\rho_{\max}$ and $e_s$ of the algorithms with and without KF.}\label{fig_3}
\end{figure}

\begin{figure}
	\centering
	\includegraphics[width = 2.5in]{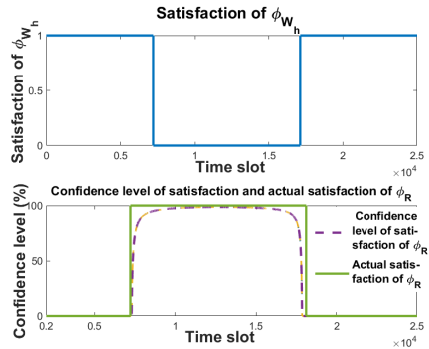}
		\vspace{-9pt}
	\caption{Progression of agents' confidence level of satisfaction of $\phi_R$, and the actual satisfaction of $\bar{s}$ over $\phi_{W_h},\phi_R$.}\label{fig_4}
\end{figure}

To show the efficacy of the KF, we also perform a centroid consensus algorithm without KF, i.e. using $\mathbf{y}$ to update $\zeta_x$ and $\zeta_y$. The actual mean estimation error is defined as $e_s = \frac{1}{N}\sum_{j=1}^N\|\boldsymbol{\zeta}^j-\bar{s}\|$. Fig. \ref{fig_3} shows the progression of $\rho_{\max}$ and $e_s$ of the distributed centroid consensus algorithm with and without KF, from which we could see $e_s$ of the consensus algorithm with KF is smaller than the one without KF, and $\rho_{\max}$ decreases as $k$ increases. 
The satisfaction of $\bar{s}$ with respect to $\phi_{W_h}$ is shown in Fig. \ref{fig_4}, where $1$ represents $\bar{s}$ satisfies $\phi_{W_h}$, and $0$ represents $\bar{s}$ violates $\phi_{W_h}$. To be more informative, we only show the satisfaction for $k\in [0,25000]$. Using the satisfaction of $\bar{s}$ over $\phi_{W_h}$, we can compute the satisfaction of $\bar{s}$ over $\phi_R$. The agents' confidence levels of $\bar{s}$ satisfying $\phi_R$ and the actual satisfaction of $\bar{s}$ over $\phi_R$ are shown in Fig. \ref{fig_4}. When $k\in[8000, 17120]$, the swarm is outside of $W_h$, so $\bar{s}$ satisfies $\phi_R$. The agents' confidence levels of satisfaction are also higher than $90\%$. In the simulation, the swarm actually stays in $W_h$ longer than $1000$ slots, so we could see $\bar{s}$ satisfies $\phi_{W_h}$ for $k\in[17120, 24000]$, and violates $\phi_R$ for $k\in[18120, 25000]$. The agents' confidence levels of satisfaction are also $0$. These results show that the agents can monitor the satisfaction of swarm features with respect to SwarmSTL formulas correctly.

\section{CONCLUSION}

In this paper, we study the problem of distributed monitoring of swarm features satisfying SwarmSTL formulas. We have developed a dynamic generalized moments consensus algorithm with Kalman filter, where each agent can estimate the generalized moments and obtain an upper bound for the estimation error. We show the estimation error bound is independent of the motion of the agents, which means the dynamic GMCA with KF can be performed simultaneously with other motion planning and control algorithms. The monitoring rules for SwarmSTL operators are proposed. The outcome of this work is that the agents can monitor whether the swarm satisfies a SwarmSTL formula with a certain confidence level. The proposed method is applied to a swarm transporting supplies example, where we also show the efficacy of the KF. Our results can be extended to synthesizing distributed controllers for a robot swarm with SwarmSTL specifications. 

\footnotesize\bibliography{reference}
\bibliographystyle{ieeetran}

\end{document}